\documentclass[aps,prl,twocolumn,superscriptaddress]{revtex4-1} 

\usepackage{graphicx}
\usepackage{epsfig}
\usepackage{bm}
\usepackage{amsmath}
\usepackage{amssymb}
\usepackage{wasysym}
\usepackage{epstopdf}
\usepackage{color}
\usepackage{xcolor}
\usepackage{epstopdf}
\usepackage{upgreek}
\usepackage{natbib}
\usepackage{arydshln}
\usepackage[T1]{fontenc}

\usepackage[colorlinks=false, bookmarks=true]{hyperref}

\begin{document}


\title{Dynamic Leidenfrost Effect: Relevant Time and Length Scales}


\author{Minori Shirota}
\altaffiliation{Both authors contributed equally on this publication.}
\affiliation{Physics of Fluids Group, Mesa+ Institute, University of Twente, 7500 AE Enschede, The Netherlands}

\author{Michiel A. J. van Limbeek}
\altaffiliation{Both authors contributed equally on this publication.}
\affiliation{Physics of Fluids Group, Mesa+ Institute, University of Twente, 7500 AE Enschede, The Netherlands}

\author{Chao Sun}
\email{Corresponding author: c.sun@utwente.nl}
\affiliation{Physics of Fluids Group, Mesa+ Institute, University of Twente, 7500 AE Enschede, The Netherlands}
\affiliation{Center for Combustion Energy and Department of Thermal Engineering, Tsinghua University, 100084 Beijing, China}

\author{Andrea Prosperetti}
\affiliation{Physics of Fluids Group, Mesa+ Institute, University of Twente, 7500 AE Enschede, The Netherlands}
\affiliation{Department of Mechanical Engineering, Johns Hopkins University, Baltimore, Maryland 21218, USA}

\author{Detlef Lohse}
\affiliation{Physics of Fluids Group, Mesa+ Institute, University of Twente, 7500 AE Enschede, The Netherlands}
\affiliation{Max Planck Institute for Dynamics and Self-Organization, 37077 Göttingen, Germany}



\date{\today}

\begin{abstract}
When a liquid droplet impacts a hot solid surface, enough vapor may be generated under it as to prevent its contact with the solid. The minimum solid temperature for this so-called Leidenfrost effect to occur is termed the Leidenfrost temperature, or the dynamic Leidenfrost temperature when the droplet velocity is non-negligible. 
We observe the wetting/drying and the levitation dynamics of the droplet 
impacting on an (isothermal) smooth sapphire surface
using high speed total internal reflection imaging, which enables us to observe the droplet base up to about 100 nm above the substrate surface. By this method we are able to reveal the processes responsible for the transitional regime between the fully wetting and the fully levitated droplet as the solid temperature increases, thus shedding light on the characteristic time- and length-scales setting the dynamic Leidenfrost temperature for droplet impact on an isothermal substrate.
\end{abstract}


\maketitle


Boiling and spreading of droplets impacting on hot substrates have been extensively studied since both phenomena strongly affect the heat transfer between the liquid and the solid. Applications include spray cooling~\cite{Kim2007},  spray combustion~\cite{Moreira2010}, and others~\cite{Attinger2000}. 

At room temperature, an impacting droplet spreads on a solid surface, and entraps a bubble under it~\cite{Bouwhuis2012,Kolinski2012,Mandre2012}. At temperatures higher than the boiling temperature $T_b$, vapor bubbles appear, which disturb and finally rupture the free surface, resulting in the violent spattering of tiny droplets~\cite{Cossali2008,Tran2012}. On even hotter surfaces, however, beyond the so-called Leidenfrost temperature $T_L$, the droplet interface becomes smooth again without any bubbles inside it. In this regime the droplet lives much longer as now it levitates on its own vapor layer: the well-known Leidenfrost effect~\cite{Chandra1991,Quere2013}.

In order to determine the Leidenfrost temperature $T_L$ and its dependence on the impact velocity $U$, 
phase diagrams have been experimentally produced for various impacting droplets with many combinations of  substrates and liquids: water on smooth silicon~\cite{Tran2012}, water on micro-structured silicon~\cite{Tran2013}, FC-72 on carbon-nanofiber~\cite{Nair2014}, water on aluminium~\cite{Bertola2015}, and ethanol on sapphire~\cite{Staat2015}.
All these phase diagrams show a weakly increasing behavior of $T_L$ with $U$.
When theoretically deriving $T_L$, one needs to determine the vapor thickness profile. In the case of a gently deposited droplet, this can be accomplished since the shape of the droplet is fixed except for the bottom surface, which reduces the problem to a  lubrication flow of vapor in the gap between the substrate and the free-surface~\cite{Wachters1966a,Snoeijer2009,Pomeau2012,Burton2012,Xu2013,Sobac2014}. 
For impacting droplets on an unheated surface at high Weber number $We \equiv \rho U^2 D_0 / \sigma$ (here $D_0$ is the 
equivalent diameter of droplet and $\rho$ and $\sigma$ are the density and the surface tension of the liquid, respectively)
it is known that the neck around the dimple beneath the impacting droplet rams the surface.
In this cold impact case, 
the neck propagates outwards like a wave  
~\cite{Mani2010}. For impact on a  superheated surface, however, it is not yet clear 
 whether  the neck still forms and rams the surface since the evaporation of the liquid in the neck and the resultant high pressure below the neck might smoothen  out the structure, resulting in a circular vapor disk with a roughly homogeneous thickness. 

\begin{figure} [t]
\begin{center}
  \includegraphics[scale=0.85]{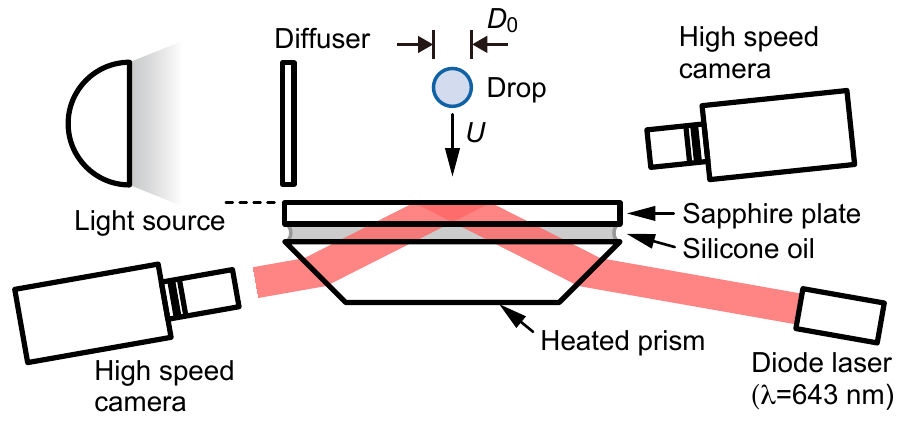}\\
\caption{Schematic of the experimental setup with synchronized side-view and TIR-imaging.\label{fgr:Setup}}
\end{center}
\end{figure}

The goal of the present paper is to experimentally clarify how the structure of the droplet base changes with increasing substrate temperature, i.e., how the characteristic time- and length-scales change at the transition  from contact to Leidenfrost boiling.
In order to explore how these scales change when undergoing the transition from contact to the Leidenfrost regime, 
we employed total internal reflection (TIR) imaging 
(see Fig.~\ref{fgr:Setup}),  
which is a powerful technique to quantitatively evaluate the approach of impacting droplets on an evanescent length scale,  typically 100 nm~\cite{Kolinski2012,Kolinski2014,Kolinski2014a}, and to clearly distinguish the wetted area from vapor bubbles/patches on heated substrates~\cite{Kim2007,Khavari2015}.

Next to the impact velocity $U$, a key process that significantly affects $T_L$ is the cooling of the substrate due to its exposure to the cold liquid.
$T_L$ thus strongly depends on the thermo-physical properties of both the liquid and the substrate used~\cite{Baumeister1973,Qiao1996,Nair2014}. 
For example, a gently deposited ethanol droplet can achieve the Leidenfrost state at $T_{L,static}$=157 $^\circ$C on polished aluminum, whereas on pyrex glass a temperature as high as 360  $^\circ$C is required. 
Water droplets, with a latent heat double that of ethanol, touch down on glass even at 700 $^\circ$C~~\cite{Baumeister1973}.
A unique wetting pattern was found on a glass substrate heated at temperatures just below $T_L$~\cite{Khavari2015}.

In this study, to avoid the complexity due to the cooling effects, we chose a combination of substrates and liquids to approximate isothermal conditions during droplet impact. We used sapphire, which has almost the same thermal properties as stainless steel, as heated surface and as liquids we used either ethanol or fluorinated heptane, the latter one of which has very low latent heat.  
With these materials, TIR-imaging allowed us to reveal the boiling characteristics at the base of impacting droplets on isothermal substrates ranging from contact boiling to Leidenfrost boiling. 

\begin{figure}
\begin{center}
\includegraphics[scale=0.5]{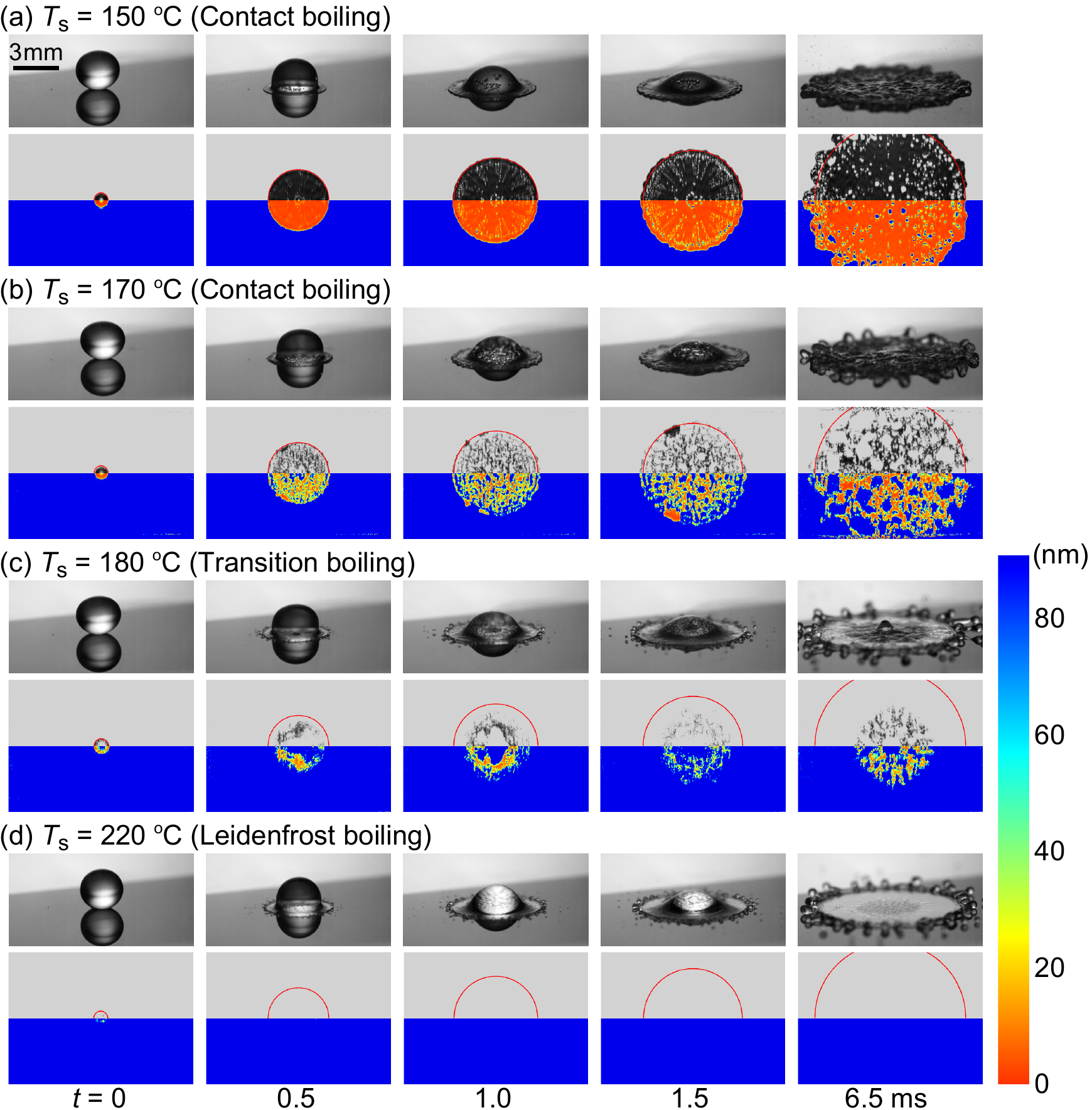}
\caption{Sequence of ethanol droplets impacting on a sapphire substrate at $T_s$ =  (a) 150, (b) 170, (c) 180 and (d) 220 $^{\circ}$C ($U$ = 1.3 m/s for all cases). The columns show images at different elapsed times after the impact; $t$ = 0, 0.5, 1.0, 1.5 and 6.5 ms from left to right. The images in the upper-row in each pair show the side view, while the lower one consists of TIR-images in original gray scale (upper part) and calculated color height scale (lower part; the image is the same as the upper one but horizontally flipped). The colored TIR-images show the distance from the substrate surface according to the height map shown in the right bottom. The cut-off height is 91 nm, and any distances more than this is shown in blue corresponding to the largest thickness. The wet area measured from the TIR-images of $T_s$ =  150 $^{\circ}$C are drawn by red half circles as a measure of the diameter of the spreading front. The corresponding movies are available as supplementary material.\label{fgr:ImSeqEth}}
\end{center}
\end{figure}

Droplets were released from a needle, which was connected to a syringe pump. We released droplets of two different liquids: ethanol and fluorinated heptane ($F_{16}C_7$), commonly known as FC-84. The liquids have almost the same boiling temperature $T_b$ ($\approx$ 80 $^{\circ}$C) but different latent heats $L$: 853 kJ/kg and 81 kJ/kg for ethanol and FC-84, respectively. The generated droplets had  a typical diameter $D_0$ of 2.8 mm for ethanol and 1.8 mm for FC-84. The impact velocity $U$ was varied by adjusting the needle height above the substrate in the range between 0.01 m and 1.5 m (measured from the droplet base), spanning the range of  $U$ = 0.4 m/s and 4.3 m/s. Both $D_0$ and $U$ were measured with a high speed camera (Photron Fastcam SA1.1) at 10 000 fps with a macro lens. For bottom view observations, we employed TIR-imaging by using a high-speed camera (Photron Fastcam SA-X2) at 40 000 fps with a long-distance microscope (Navitar 12x Telecentric zoom system).
Both side and bottom view images provided fields of view of about 10 $\times$ 10 mm$^2$, and spatial resolution of about 20 $\mu$m/pixel.

The droplets impacted on a smooth sapphire substrate (50 mm in diameter and 3 mm in thickness) with a roughness of 10 nm (measured by AFM).
The substrate was placed on a glass dove prism with a high-viscous silicone oil (kinematic viscosity: 0.01 $m^2/s$) between them for optical impedance matching (Fig.~\ref{fgr:Setup}). The prism was mounted in an aluminium heating block whose temperature was PID-controlled to a fixed value ranging from 80 $^{\circ}$C to 590 $^{\circ}$C by two electrical heating cartridges and a thermal probe. 
The exact temperature on the substrate surface was measured before the experiment with a surface probe. 
\begin{figure*}[t]
\begin{center}
  \includegraphics[scale=1.0]{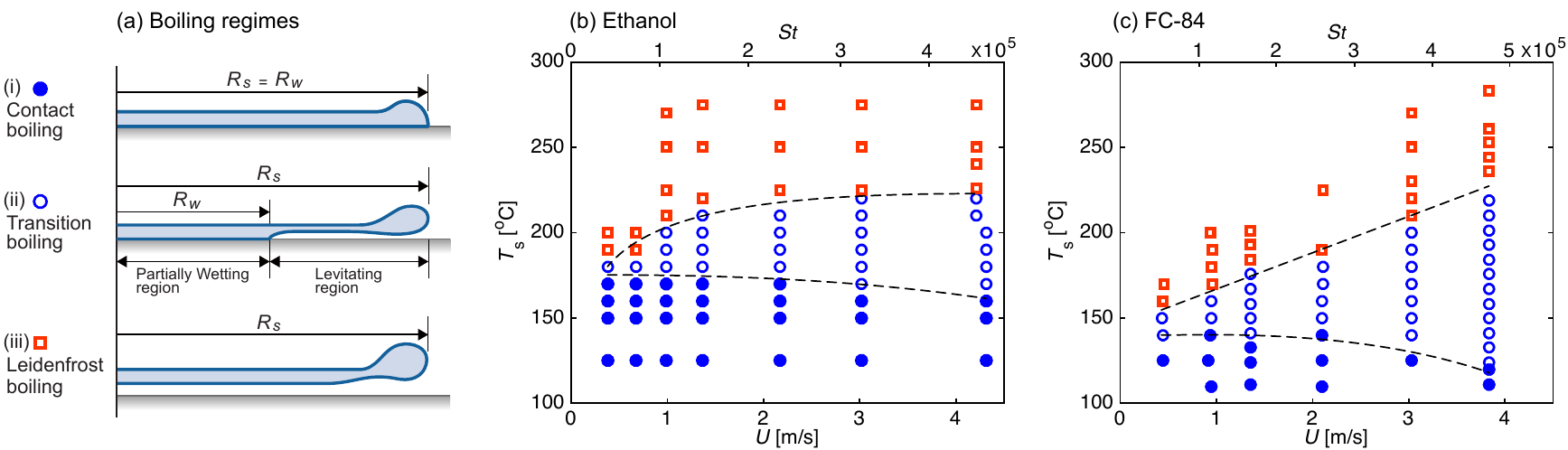}\\
  \caption{(a) Schematic diagrams of the three different boiling regimes identified in the present study, (i) contact, (ii) transition, and (iii) Leidenfrost, showing the differences between the radius of the spreading front $R_s$ and that one of the (partially) wetting region $R_w$. Phase diagram of the boiling regimes for (b) ethanol and (c) FC-84 with its more than ten times lower latent heat (as compared to ethanol). The plots indicate the three different boiling behaviors: the contact boiling regime (solid circle), the transition regime (open circle), and the Leidenfrost regime (square). The dashed lines between different regimes are drawn to guide the eye. For comparison, static Leidenfrost temperatures are 160~$^\circ$C and 125~$^\circ$C for ethanol and FC-84, respectively. \label{fgr:PH}}
\end{center}
\end{figure*}

For TIR-imaging, a diode laser beam (wave length: 643 nm) was expanded to about 20 mm in diameter, and introduced to the dove prism via mirrors at a certain incident angle. 
Since the intensity of an evanescent light exponentially decays with a distance from the substrate~\cite{Hecht2002}, the logarithmic intensity of the droplet image normalized by the one without the droplet is proportional to the distance. The proportionality is determined by the wave length and the incident angle of the laser, and the reflective indices of the substrate and the gas above the substrate. When the droplet touches the substrate, the corresponding part of laser light transmits through the droplet, and therefore we can  clearly distinguish the wetted area from the dry one as a sharp change in gray-scale intensity. The resulting image of the contacting droplet is not a circle but an ellipsoid since the image shrinks only in the direction along the side wall of the prism with the oblateness according to the incident angle. From the oblateness of the droplet image, therefore, we can calculate the incident angle which is the key parameter to quantitatively evaluate the decaying length of an evanescent wave (see supplementary material). The relative uncertainty is found to be 10 \% for the ethanol and 7 \% for the FC-84 measurements. 

We observed in detail the behavior of different boiling regimes with both side view and TIR images for ethanol. 
Figure~\ref{fgr:ImSeqEth} shows the side view and TIR-images. The latter consist of original gray scale images (upper part) and color height images (lower part).
Just above the boiling temperature $T_b$, from the color height images, we can clearly see that the droplet completely wets the substrate except for the area of the nucleated bubbles, see Fig.~\ref{fgr:ImSeqEth} (a). 
With increasing substrate temperature $T_s$, the growth and coalescence of the nucleated bubbles are enhanced (Fig.~\ref{fgr:ImSeqEth} (b)). 
Although at \textit{t} = 0 the droplet is in contact with the substrate aside from the central dimple region, 
only partial contact occurs later as indicated by smaller red areas and larger green areas in comparison with the lower temperature case. 
The color height images indicate the corrugated surface of the droplet base. However, the droplet surface is still within the length scale of evanescent light, 90 nm in this case.
In addition, the radius of the periphery of the wetted area (red circle) is almost the same as that for $T_s$ = 150 $^{\circ}$C (red circle). We thus categorize this regime still as contact boiling. 

A further increase of the substrate temperature drastically changes the boiling behavior, as sketched by the difference between Fig.~\ref{fgr:PH} (a-i) and (a-ii). This regime, to which we refer as transition boiling, sets in at $T_s$ = 180 $^{\circ}$C (Fig.~\ref{fgr:ImSeqEth} (c)). The TIR images reveal that the wetted area is smaller than the spreading radius of the droplet, as shown by the wetted area being smaller than the red circle. The local evaporation in the outer area of the base causes the levitation of the lamella which becomes unstable and breaks up as shown in the side view images. 
The onset of the rim instability in the early stage of the impact (at $t$ = 0.5 ms in Fig.~\ref{fgr:ImSeqEth} (c) and (d)) is thus a good indicator for the end of the contact boiling regime. 
A close observation of the change in color height maps reveals that although most of the contact area is not wetted at $t$ = 1.5 ms, some contact (red area) is recovered later at $t$ = 6.5 ms. This can be explained by the cooling by the vapor from the droplet base located at about 50 nm from the substrate. We also found that there exists no fingering pattern in contrast to what was observed on glass substrates~\cite{Khavari2015}. These results indicate that the fingering pattern is related to the cooling of the substrate by radially flowing vapor and subsequent re-wetting of the substrate. 

At higher temperatures, the droplet never touches the substrate during the impact process, either temporarily or spatially: the Leidenfrost boiling regime sets in (Fig.~\ref{fgr:ImSeqEth}(d)). 
In the side view images, we can also see that the levitation on the vapor layer changes the brightness of the droplet  (cf. the four side view images at $t$ = 1.0 ms in Fig.~\ref{fgr:ImSeqEth}): here the backlight used for the side view imaging undergoes a total internal reflection at the top surface of the vapor layer, resulting in the bright color of the droplet in the Leidenfrost state. 

In summary, on our nearly isothermal substrates, we have identified three different boiling regimes, i.e., contact, transition, and Leidenfrost (Fig.~\ref{fgr:PH}~(a)).  In contact boiling, the radius $R_s$ of the spreading front  coincides with that of (partially) wetted region $R_w$, whereas in the transition boiling, $R_w $ is smaller than $R_s$. In the Leidenfrost regime, $R_w=0$ during the whole spreading process, i.e., no wetting at all. 
Note that the present classification of the boiling regimes
 is based on the {\it direct observation}  of wet respective dry areas, and is thus 
 different from the classification 
  in  previous studies~\cite{Tran2012,Tran2013} where the boiling regimes were more superficially classified based on the smoothness of the droplet surface or the ejection of 
   tiny droplets from the impacting droplet.    
On the basis of the definitions described above, the boiling regimes can be classified into contact, transition and Leidenfrost for impact velocities ranging from about 0.5 to 4 m/s for both liquids. The phase diagrams of Fig.~\ref{fgr:PH} (b) and (c) show a considerable temperature range for transition boiling, approximately 50 K and 100 K for ethanol and FC-84, respectively. 

For ethanol, the lower boundary between contact and transition boiling depends very weakly on $U$, if at all. The higher boundary between transition and film boiling, i.e. $T_L$, increases with $U$ only weakly. In addition, our experiments with a sapphire substrate show a $T_L$ lower by about 50 K  as compared with glass~\cite{Khavari2015}.  
This finding confirms that significant heat is lost from the substrate by cooling even during the short impact time. 
\begin{figure}[t]
\begin{center}
\includegraphics[scale = 1.0]{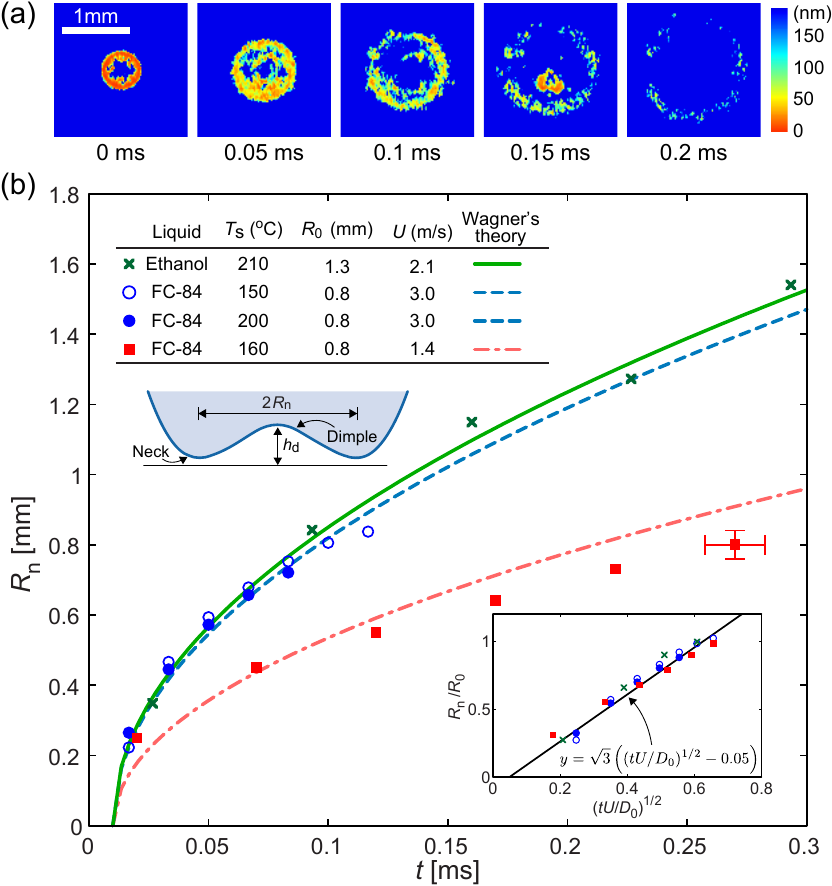}
\caption{Spreading of a circular liquid ring within the evanescent length scale at the beginning of the impact. 
(a) Color height images obtained from TIR-images of the droplet base, for FC-84, $T_s$ = 160 $^{\circ}$C and $U$ = 1.4 m/s. 
(b) Radius of the circular ring, $R_n$, corresponding to the closest area of the droplet to the substrate. The error bars represent the uncertainty due to the frame interval in the imaging (horizontal) and the pixel resolution of the image (vertical). The curves show the position of the maximum pressure area, estimated by a modified Wagner's theory: $\sqrt{3R_0Ut}$. The inset shows the same data but in dimensionless form. The data collapse reveals the universality.\label{fgr:WagRing}}
\end{center}
\end{figure}

For FC-84, the contact-to-transition boundary is lower than that for ethanol due to its mote than ten times lower latent heat as compared to ethanol. At low impact velocities, the upper $T_L$ boundary is also lower than for ethanol, but it increases with $U$ and reaches almost the same level as that for ethanol at $U \approx$ 4 m/s.
Several 
 models for the spreading diameter $D_{max}$ of an impacting drop have been developed~\cite{Chandra1991,Clanet2004,Tran2013,Lastakowski2014}.
For the so-called pancake model \cite{Tran2013}
 $D_{max}$ and $D_0/U$ are considered to be  the relevant
  length- and time-scales~\cite{Tran2013}.  In addition, the model assumes a roughly 
  homogeneous thickness for the vapor layer below the droplet. 
However, this model, combined with a force balance on the levitating droplet base (shear stress of vapor $\sim$ capillary pressure), does not 
correctly describe the dependence of $T_L$ on $U$,
 because it predicts a scaling $T_L - T_b \sim U^{-2}$, which is  neither consistent with the current experimental results (Fig.~\ref{fgr:PH} (b) and (c)) 
 nor with previous ones~\cite{Tran2012,Tran2013,Nair2014,Bertola2015}. 

Our TIR-images shown in Fig.~\ref{fgr:WagRing} (a) reveal the crucial deficiencies of the pancake model: just below $T_L$, 
 in the very beginning of the impact process ($tU/D<$ 0.1)
a circular ring of radius $R_n(t)$ that is  {\it smaller}  than the undisturbed droplet radius ($R_n/R_0 < 1$) becomes
 visible within the evanescent length scale. 
Based on this  finding
 as well as on  the fact that the ring corresponds to the  area of the droplet base closest to the substrate, we 
 conclude 
  that the ring shows the position of the neck around the central dimple, see the sketch shown as an inset of Fig.~\ref{fgr:WagRing} (b).
The pancake model is therefore inappropriate 
 for the modelling of $T_L$ because of its much larger time- and length-scales and the erroneous 
 assumption of a roughly  homogeneous vapor film thickness.   

In their study of a droplet impacting on a cold surface, Riboux and Gordillo~\cite{Riboux2014} have pointed out the similarity with the classical problem of a solid body impacting a liquid surface first treated by Wagner~\cite{Wagner1932} (see also~\cite{Howison1991,Zhao1993}).
They show the relevance of Wagner's prediction $\sqrt{3R_0Ut}$ for the radius of the liquid-solid contact region for the droplet impact problem.
The situation in the present case is different because the formation of the dimple and the associated neck precedes the actual liquid-solid contact.
However, a strong similarity can be found in the very high pressure that develops under the droplet due to the vapor formation.
This consideration has prompted us to examine the possible connection between the neck radius $R_n$ on the lower surface of the droplet and Wagner's result. 

The symbols in Fig.~\ref{fgr:WagRing} (b) are the measured position of the neck around the central dimple for four cases, while the lines show Wagner's result $\sqrt{3R_0Ut}$.
The striking agreement for the two liquids, and various $T_s$, $R_0$, and $U$ (Fig. ~\ref{fgr:WagRing} (b)) suggests that the concept introduced by Riboux and Gordillo~\cite{Riboux2014} for impact on a cold substrate is equally applicable to the present hot substrate case.
This finding reveals that 
the relevant time- and length-scales for the droplet touch-down on a superheated solid surface are the
dimple formation time $t_d \sim h_d/U$ and the radius of the neck $R_n$.
These  scale as 
$t_d U/R_0 \sim h_d/R_0 \sim St^{-2/3}$ and 
 $R_n/R_0 \sim \sqrt{3Ut_d/R_0} \sim St^{-1/3}$, where we used the well-known scaling for the dimple height (see the inset of Fig. 4(b)) $h_d \sim R_0 St^{-2/3}$ in the inertia-dominant regime~\cite{Mani2010,Bouwhuis2012} with the  Stokes number ${St= \rho R_0 U/\mu_v}$ based on the vapor viscosity $\mu_v$.  

In conclusion, the use of total internal reflection (TIR) imaging has permitted us to resolve 
the processes occurring under an impacting liquid droplet within 100 nm above an isothermal superheated surface.
We for the first time quantitatively evaluated the height of the droplet base, levitating on its vapor within the evanescence length scale in the transition regime between the contact and the Leidenfrost regimes. 
The data are presented in the form of phase diagrams in the temperature-velocity parameter space. 
At temperatures just below the Leidenfrost threshold $T_L$, a circular liquid ring is seen to spread within the evanescent length scale at the beginning of the impact process.   
This observation provided us with the crucial ingredient for future modelling of $T_L$, namely the importance of the dimple and neck structures under the droplet.


\begin{acknowledgments}
We thank Sander Wildeman for illuminating discussion.
This work was partially supported by Fundamenteel Onderzoek der Materie and by an ERC-Advanced Grant.
\end{acknowledgments}


\bibliography{PRL_Ref_abbrv.bib}

\end{document}